%% file: ad2016a.tex
\documentclass[10pt,a4paper]{article} 
\usepackage[margin=1.5cm]{geometry}
\usepackage[numbers]{natbib}
\usepackage{amsmath}
\usepackage{paralist}
\usepackage[hidelinks]{hyperref}
\hypersetup{colorlinks,linkcolor={blue},citecolor={blue},urlcolor={blue}}
\usepackage{MnSymbol}
\usepackage{graphicx}
\DeclareGraphicsExtensions{.pdf}
\usepackage{listings}
\usepackage{color}
\usepackage{wrapfig}
\include{AD}

\title{Binomial Checkpointing for Arbitrary Programs with No User Annotation%
\thanks{\textbf{Extended abstract presented at the AD 2016 Conference, Sep 2016, Oxford UK.}}}
\author{\href{http://engineering.purdue.edu/~qobi}{Jeffrey Mark
    Siskind}\footnote{Corresponding Author, School of Electrical and Computer
    Engineering, Purdue University,
    \href{mailto:qobi@purdue.edu}{\texttt{qobi@purdue.edu}}}
  \qquad
  \href{http://barak.pearlmutter.net}{Barak A.
    Pearlmutter}\footnote{Dept of Computer Science, National University of
    Ireland Maynooth,
    \href{mailto:barak@pearlmutter.net}{\texttt{barak@pearlmutter.net}}}}
\date{April 2016}

\newcommand{\CHECKPOINT}[4]{\lsem#1,#2,#3,#4\rsem}
\newcommand{\CHECKPOINTREVERSEJ}{\begin{array}[b]{@{}l@{}}
\!\text{\small$\checkmark$}\\*[-6pt]
\mathcal{J}\end{array}}

\newcommand{\eg}{\emph{e.g.},}

\newcommand{\ie}{\emph{i.e.},}

\newcommand{\Tapenade}{\textsc{Tapenade}}
\newcommand{\SMLNJ}{\textsc{sml/nj}}
\newcommand{\SML}{\textsc{sml}}
\newcommand{\MLton}{\textsc{MLton}}
\newcommand{\Clang}{\textsc{c}}
\newcommand{\POSIX}{\textsc{posix}}
\newcommand{\Fortran}{\textsc{Fortran}}
\newcommand{\Prolog}{\textsc{Prolog}}
\newcommand{\ADOLC}{\textsc{adol-c}}
\newcommand{\VLAD}{\textsc{vlad}}
\newcommand{\FADBAD}{\textsc{fadbad}$++$}
\newcommand{\Stalingrad}{\textsc{Stalin}$\nabla$}

\definecolor{darkblue}{rgb}{0,0,0.7}
\definecolor{darkgreen}{rgb}{0,0.5,0}
\definecolor{darkred}{rgb}{0.7,0,0}

\begin{document}
\maketitle
\thispagestyle{empty}

Heretofore, automatic checkpointing at procedure-call boundaries
\citep{Volin1985Aco}, to reduce the space complexity of reverse mode, has been
provided by systems like \Tapenade\ \citep{Dauvergne2006TDF}.
However, binomial checkpointing, or treeverse \citep{Griewank1992ALG}, has only
been provided in AD systems in special cases, \eg\ through user-provided
pragmas on DO loops in \Tapenade, or as the nested taping mechanism in
\ADOLC\ for time integration processes, which requires that user code be
refactored.
We present a framework for applying binomial checkpointing to arbitrary code
with no special annotation or refactoring required.
This is accomplished by applying binomial checkpointing directly to a program
trace.
This trace is produced by a general-purpose checkpointing mechanism that is
orthogonal to AD.\@

\begin{wrapfigure}[10]{r}[-0.35em]{35ex}
  \par\vspace*{-5.5ex}
\begin{lstlisting}[frame=single,caption={\Fortran\ example},label=lis:fortran]
      subroutine f(x, y)
      n = 100003
      y = x
c$ad binomial-ckp n+1 30 1
      do i = 1, n
         m = l(x, i)
         do j = 1, m
            y = y*y
            y = sqrt(y)
         end do
      end do
      end
\end{lstlisting}
  \par\vspace*{-2ex}
\end{wrapfigure}
Consider the code fragment in Listing~\ref{lis:fortran}.
This example, $y=f(x)$, while contrived, is a simple caricature of a situation
that arises commonly in practice, \eg\ in adaptive grid methods.
Here, the duration of the inner loop varies wildly as some function~$l(x, i)$
of the input and the outer loop index, perhaps
$2^{\lfloor\lg(n)\rfloor-\lfloor\lg(1+(1007\lfloor3^x\rfloor i\mod n))\rfloor}$,
that is small on most iterations of the outer loop but $O(n)$ on a few
iterations.
Thus the optimality of the binomial schedule is violated.
The issue is that the optimality of the binomial schedule holds at the level of
primitive atomic computations but this is not reflected in the static syntactic
structure of the source code.
Often, the user is unaware or even unconcerned with the micro-level structure
of atomic computations and does not wish to break the modularity of the source
code to expose such.
Yet the user may still wish to reap the benefits of an optimal binomial
checkpointing schedule \citep{Griewank2000ARA}.
Moreover, the relative duration of different paths through a program may vary
from loop iteration to loop iteration in a fashion that is data dependent, as
shown by the above example, and not even statically determinable.
We present an implementation strategy for checkpointing that does not require
user placement of checkpoints and does not constrain checkpoints to subroutine
boundaries, DO loops, or other syntactic program constructs.
Instead, it can automatically and dynamically introduce a checkpoint at an
arbitrary point in the computation that need not correspond to a syntactic
program unit.

%
\begin{wrapfigure}[19]{r}{314pt}
  \par\vspace*{-1.5ex}
  \fbox{
    \begin{math}
      \begin{aligned}
        \APPLY\;\CLOSURE{\lambda x.e}{\rho}\;v&=\EVAL\;\rho[x\mapsto v]\;e\\
        \FORWARDJ\;v_1\;v_2\;\acute{v}_3&=
        \LET(\BUNDLE{v_4}{\acute{v}_5})=\APPLY\;v_1\;(\BUNDLE{v_2}{\acute{v}_3})
        \;\IN(v_4,\acute{v}_5)\\
        \REVERSEJ\;v_1\;v_2\;\grave{v}_3&=
        \LET(\TAPE{v_4}{\grave{v}_5})=(\TAPE{(\APPLY\;v_1\;v_2)}{\grave{v}_3})
        \;\IN(v_4,\grave{v}_5)\\
        \EVAL\;\rho\;c&=c\\
        \EVAL\;\rho\;x&=\rho\;x\\
        \EVAL\;\rho\;(\lambda x.e)&=\CLOSURE{\lambda x.e}{\rho}\\
        \EVAL\;\rho\;(e_1\;e_2)&=
        \APPLY\;(\EVAL\;\rho\;e_1)\;(\EVAL\;\rho\;e_2)\\
        \EVAL\;\rho\;(\IF e_1\;\THEN e_2\;\ELSE e_3)&=
        \IF(\EVAL\;\rho\;e_1)\;
        \THEN(\EVAL\;\rho\;e_2)\;
        \ELSE(\EVAL\;\rho\;e_3)\\
        \EVAL\;\rho\;(\diamond e)&=\diamond(\EVAL\;\rho\;e)\\
        \EVAL\;\rho\;(e_1\bullet e_2)&=
        (\EVAL\;\rho\;e_1)\bullet(\EVAL\;\rho\;e_2)\\
        \EVAL\;\rho\;(\FORWARDJ\;e_1\;e_2\;e_3)&=
        \FORWARDJ\;(\EVAL\;\rho\;e_1)\;(\EVAL\;\rho\;e_2)\;(\EVAL\;\rho\;e_3)\\
        \EVAL\;\rho\;(\REVERSEJ\;e_1\;e_2\;e_3)&=
        \REVERSEJ\;(\EVAL\;\rho\;e_1)\;(\EVAL\;\rho\;e_2)\;(\EVAL\;\rho\;e_3)
      \end{aligned}
    \end{math}}
  \par\vspace*{-1.5ex}
  \caption{Direct-style evaluator for \VLAD.}
  \label{fig:direct}
\end{wrapfigure}
%
We have previously introduced \VLAD, a pure functional language with builtin AD
operators for both forward and reverse mode.
Here, we adopt slight variants of these operators with the following signatures.
\begin{equation*}
  \FORWARDJ:f\;x\;\acute{x}\mapsto y\;\acute{y}\hspace*{2em}
  \REVERSEJ:f\;x\;\grave{y}\mapsto y\;\grave{x}
\end{equation*}
The~$\FORWARDJ$ operator calls a function~$f$ on a primal~$x$ with a
tangent~$\acute{x}$ to yield a primal~$y$ and a tangent~$\acute{y}$.
The~$\REVERSEJ$ operator calls a function~$f$ on a primal~$x$ with a
cotangent~$\grave{y}$ to yield a primal~$y$ and a cotangent~$\acute{x}$.
Here, we restrict ourselves to the case where (co)tangents are ground data
values, \ie\ reals and (arbitrary) data structures containing reals and other
scalar values, but not functions (\ie\ closures).
For our purposes, the crucial aspect of the design is that the AD operators are
provided within the language, since these provide the portal to the
checkpointing mechanism.

In previous work, we introduced \Stalingrad, a highly optimizing compiler for
\VLAD.
Here, we formulate a simple evaluator (interpreter) for
\VLAD\ (Fig.~\ref{fig:direct}) and extend such to perform binomial
checkpointing.
The operators~$\diamond$ and~$\bullet$ range over the unary and binary basis
functions respectively.
This evaluator is written in what is known in the programming-language
community as \defoccur{direct style}, where functions (in this case~$\EVAL$,
denoting `eval', $\APPLY$, denoting `apply', and the implementations
of~$\FORWARDJ$ and~$\REVERSEJ$ in the host) take inputs as function-call
arguments and yield outputs as function-call return values
\citep{reynolds1993discoveries}.
AD is performed by overloading the basis functions in the host, in a fashion
similar to \FADBAD\ \citep{Bendtsen1996FaF},
$\BUNDLE{x}{\acute{x}}$ denotes recursively bundling a data structure
containing primals with a data structure containing tangents, or alternatively
recursively unbundling such when used as a binder, and $\TAPE{y}{\grave{y}}$
denotes running the reverse sweep on the tape~$y$ with the output
cotangent~$\grave{y}$, or alternatively extracting the primal~$y$ and input
cotangent~$\grave{x}$ from the tape when used as a binder $\TAPE{y}{\grave{x}}$.

We introduce a new AD operator~$\CHECKPOINTREVERSEJ$ to perform binomial
checkpointing.
The crucial aspect of the design is that the signature (and semantics)
of~$\CHECKPOINTREVERSEJ$ is \emph{identical} to~$\REVERSEJ$; they are
\emph{completely interchangeable}, differing only in the space/time complexity
tradeoffs.
This means that code \emph{need not be modified} to switch back and forth
between ordinary reverse mode and binomial checkpointing, save interchanging
calls to~$\REVERSEJ$ and~$\CHECKPOINTREVERSEJ$.

\begin{wrapfigure}[9]{r}{235pt}
  \vspace*{-5ex}
  \fbox{
    \begin{tabular}{lll@{\hspace{2em}}l@{}}
      \multicolumn{4}{l}{To compute $(y,\grave{x})=
        \CHECKPOINTREVERSEJ\;f\;x\;\grave{y}$:}\\
      & \textbf{base case} ($f\;x$ fast):
      & $(y,\grave{x})=\REVERSEJ\;f\;x\;\grave{y}$ & (0)\\[1ex]
      & \textbf{inductive case}:
      & $h\circ g=f$ & (1)\\
      && $u=g\;x$ & (2)\\
      && $(y,\grave{u})=\CHECKPOINTREVERSEJ\;h\;u\;\grave{y}$ & (3)\\
      && $(u,\grave{x})=\CHECKPOINTREVERSEJ\;g\;x\;\grave{u}$ & (4)
    \end{tabular}}
  \par\vspace*{-1.5ex}
  \caption{Algorithm for binomial checkpointing.}
  \label{fig:binomial}
\end{wrapfigure}
Conceptually, the behavior of~$\CHECKPOINTREVERSEJ$ is shown in
Fig.~\ref{fig:binomial}.
In this inductive definition, a function~$f$ is split into the composition of
two functions~$g$ and~$h$ in step~1, the checkpoint~$u$ is computed by
applying~$g$ to the input~$x$ in step~2, and the cotangent is computed by
recursively applying~$\CHECKPOINTREVERSEJ$ to~$h$ and~$g$ in steps~3 and~4.
This divide-and-conquer behavior is terminated in a base case, when the
function~$f$ is small, at which point the cotangent is computed
with~$\REVERSEJ$, in step~0.
If step~1 splits a function~$f$ into two functions~$g$ and~$h$ that take the
same number of computational steps, the recursive divide-and-conquer process
yields the logarithmic asymptotic space/time complexity of binomial
checkpointing.

The central difficulty in implementing the above is performing step~1, namely
splitting a function~$f$ into two functions~$g$ and~$h$, ideally ones that take
the same number of computational steps.
A sophisticated user can manually rewrite a subroutine~$f$ into two
subroutines~$g$ and~$h$.
A sufficiently powerful compiler or source transformation tool might also be
able to, with access to nonlocal program text.
But an overloading system, with access only to local information, would not be
able to.

We solve this problem by providing an interface to a general-purpose
checkpointing mechanism orthogonal to AD.\@
\begin{center}
  \begin{tabular}{ll}
    $\textsc{primops}\;f\;x\mapsto(y,n)$&Return $y=f(x)$ along with the number
    $n$ of steps needed to compute $y$.\\
    $\textsc{checkpoint}\;f\;x\;n\mapsto u$&Run the first $n$ steps of the
    computation of $f(x)$ and return a checkpoint $u$.\\
    $\textsc{resume}\;u\mapsto y$&If $u=(\textsc{checkpoint}\;f\;x\;n)$, return
    $y=f(x)$.
  \end{tabular}
\end{center}
This interface allows (a)~determining the number of steps of a computation,
(b)~interrupting a computation after a specified number of steps, usually half
the number of steps determined by the mechanism in~(a), and (c)~resuming an
interrupted computation to completion.
A variety of implementation strategies for this interface are possible.
We present one in detail momentarily and briefly discuss others below.

\begin{wrapfigure}[9]{r}{315pt}
  \par\vspace*{-3.5ex}
  \fbox{
    \begin{tabular}{lll@{\hspace{2em}}l@{}}
      \multicolumn{4}{l}{To compute $(y,\grave{x})=
        \CHECKPOINTREVERSEJ\;f\;x\;\grave{y}$:}\\
      & \textbf{base case:}
      & $(y,\grave{x})=\REVERSEJ\;f\;x\;\grave{y}$ & (0)\\
      & \textbf{inductive case:}
      & $(y,2n)=\textsc{primops}\;f\;x$ & (1)\\
      && $u=\textsc{checkpoint}\;f\;x\;n$ & (2)\\
      && $(y,\grave{u})=
      \CHECKPOINTREVERSEJ\;(\lambda u.\textsc{resume}\;u)\;u\;\grave{y}$ & (3)\\
      && $(u,\grave{x})=
      \CHECKPOINTREVERSEJ\;
      (\lambda x.\textsc{checkpoint}\;f\;x\;n) \;x\;\grave{u}$ & (4)
  \end{tabular}}
  \par\vspace*{-2ex}
  \caption{Binomial checkpointing via general checkpointing interface.}
  \label{fig:implementation}
\end{wrapfigure}
Irrespective of how one implements the general-purpose checkpointing interface,
one can use it to implement~$\CHECKPOINTREVERSEJ$ as shown in
Fig.~\ref{fig:implementation}.
The function~$f$ is split into the composition of two
functions~$g$ and~$h$ by taking~$g$ as $\lambda x.\textsc{checkpoint}\;f\;x\;n$,
where~$n$ is half the number of steps determined by $\textsc{primops}\;f\;x$,
and~$h$ as $\lambda u.\textsc{resume}\;u$.

\setcounter{figure}{4}
\begin{figure}
  \resizebox{\textwidth}{!}{\fbox{
  \begin{math}
  \begin{aligned}
    \APPLY\;k\;n\;l\;\CLOSURE{\lambda x.e}{\rho}\;v&=
    \EVAL\;k\;n\;l\;\rho[x\mapsto v]\;e\\
    \FORWARDJ\;k\;n\;l\;v_1\;v_2\;\acute{v}_3&=
    \APPLY\;(\lambda n\;l\;(\BUNDLE{v_4}{\acute{v}_5}).
    k\;n\;l\;(v_4,\acute{v}_5))
    \;n\;l\;v_1\;(\BUNDLE{v_2}{\acute{v}_3})\\
    \REVERSEJ\;k\;n\;l\;v_1\;v_2\;\acute{v}_3&=
    \APPLY\;(\lambda n\;l\;v.
    \LET(\TAPE{v_4}{\grave{v}_5})=\TAPE{v}{\grave{v}_3}
    \;\IN k\;n\;l\;(v_4,\acute{v}_5))
    \;n\;l\;v_1\;v_2\\
    \EVAL\;k\;l\;l\;\rho\;e&=\CHECKPOINT{k}{l}{\rho}{e}\\
    \EVAL\;k\;n\;l\;\rho\;c&=k\;(n+1)\;l\;c\\
    \EVAL\;k\;n\;l\;\rho\;x&=k\;(n+1)\;l\;(\rho\;x)\\
    \EVAL\;k\;n\;l\;\rho\;(\lambda x.e)&=
    k\;(n+1)\;l\;\CLOSURE{\lambda x.e}{\rho}\\
    \EVAL\;k\;n\;l\;\rho\;(e_1\;e_2)&=
    \EVAL\;(\lambda n\;l\;v_1.
    (\EVAL\;(\lambda n\;l\;v_2.
    (\APPLY\;k\;n\;l\;v_1\;v_2))
    \;n\;l\;\rho\;e_2))
    \;(n+1)\;l\;\rho\;e_1\\
    \EVAL\;k\;n\;l\;\rho\;(\IF e_1\;\THEN e_2\;\ELSE e_3)&=
    \EVAL\;(\lambda n\;l\;v_1.(\IF v_1\;
    \THEN(\EVAL\;k\;n\;l\;\rho\;e_2)\;
    \ELSE(\EVAL\;k\;n\;l\;\rho\;e_3)))\;(n+1)\;l\;\rho\;e_1\\
    \EVAL\;k\;n\;l\;\rho\;(\diamond e)&=
    \EVAL\;(\lambda n\;l\;v.(k\;n\;l\;(\diamond v)))\;(n+1)\;l\;\rho\;e\\
    \EVAL\;k\;n\;l\;\rho\;(e_1\bullet e_2)&=
    \EVAL\;(\lambda n\;l\;v_1.
    (\EVAL\;(\lambda n\;l\;v_2.
    (k\;n\;l\;(v_1\bullet v_2)))
    \;n\;l\;\rho\;e_2))
    \;(n+1)\;l\;\rho\;e_1\\
    \EVAL\;k\;n\;l\;\rho\;(\FORWARDJ\;e_1\;e_2\;e_3)&=
    \EVAL\;(\lambda n\;l\;v_1.
    (\EVAL\;(\lambda n\;l\;v_2.
    (\EVAL\;(\lambda n\;l\;v_3.
    (\FORWARDJ\;k\;n\;l\;v_1\;v_2\;v_3))
    \;n\;l\;\rho\;e_3))
    \;n\;l\;\rho\;e_2))
    (n+1)\;l\;\;\rho\;e_1\\
    \EVAL\;k\;n\;l\;\rho\;(\REVERSEJ\;e_1\;e_2\;e_3)&=
    \EVAL\;(\lambda n\;l\;v_1.
    (\EVAL\;(\lambda n\;l\;v_2.
    (\EVAL\;(\lambda n\;l\;v_3.
    (\REVERSEJ\;k\;n\;l\;v_1\;v_2\;v_3))
    \;n\;l\;\rho\;e_3))
    \;n\;l\;\rho\;e_2))
    (n+1)\;l\;\rho\;e_1
  \end{aligned}
  \end{math}}}
  \par\vspace*{-1ex}
  \caption{CPS evaluator for \VLAD.}
  \label{fig:cps}
  \par\vspace*{-3ex}
\end{figure}

One way of implementing the general-purpose checkpointing interface is to
convert the evaluator from direct style to continuation-passing style (CPS,
\citep{sussman1975}), where functions (in this case~$\EVAL$, $\APPLY$,
$\FORWARDJ$, and~$\REVERSEJ$ in the host) take an additional continuation
input~$k$ and instead of yielding outputs via function-call return, do so by
calling the continuation with said output as arguments (Fig.~\ref{fig:cps}).
In such a style, functions never return; they just call their continuation.
With tail-call merging, such corresponds to a computed \texttt{go to} and does
not incur stack growth.
This crucially allows the interruption process to actually return a checkpoint
data structure containing the saved state of the evaluator, including its
continuation, allowing the evaluation to be resumed by calling the evaluator
with this saved state.
This `level shift' of return to calling a continuation allowing an actual
return to constitute checkpointing interruption is analogous to the way
backtracking is classically implemented in \Prolog, with success implemented as
calling a continuation and failure implemented as actual return.
In our case, we further instrument the evaluator to thread two values as inputs
and outputs: the count~$n$ of the number of evaluation steps, which is
incremented at each call to~$\EVAL$, and the limit~$l$ of the number of steps,
after which a checkpointing interrupt is triggered.

\setcounter{figure}{3}
\begin{wrapfigure}[6]{r}{217pt}
  \par\vspace*{-4ex}
  \fbox{
    \begin{math}
      \begin{aligned}
        \textsc{primops}\;f\;x&=
        \APPLY\;(\lambda n\;l\;v.(v,n)))\;0\;\infty\;f\;x\\
        \textsc{checkpoint}\;f\;x\;n&=\APPLY\;\bot\;0\;n\;f\;x\\
        \textsc{resume}\;\CHECKPOINT{k}{l}{\rho}{e}&=
        \EVAL\;k\;l\;\infty\;\rho\;e\\
      \end{aligned}
  \end{math}}
  \par\vspace*{-2ex}
  \caption{Implementation of the general-purpose checkpointing interface using
    the CPS evaluator.}
  \label{fig:interface}
\end{wrapfigure}
With this CPS evaluator, it is possible to implement the general-purpose
checkpointing interface (Fig.~\ref{fig:interface}), not for programs in the
host, but for programs in the target; hence our choice of formulating the
implementation around an evaluator (interpreter).
We remove this restriction below.
The implementation of \textsc{primops} calls the evaluator with no limit and
simply counts the number of steps to completion.
The implementation of \textsc{checkpoint} calls the evaluator with a limit that
must be smaller than that needed to complete so a checkpointing interrupt is
forced and the checkpoint data structure $\CHECKPOINT{k}{l}{\rho}{e}$ is
returned.
The implementation of \textsc{resume} calls the evaluator with arguments from
the saved checkpoint data structure.

\lstset{language=Lisp,
        morekeywords={begin,either},
        deletekeywords={remove},
	basicstyle=\ttfamily\small,
	keywordstyle=\color{darkblue}\bfseries\ttfamily\small,
	commentstyle=\color{darkred}\itshape\ttfamily\small}

\begin{wrapfigure}[10]{r}[-0.3em]{4in}
  \par\vspace*{-5ex}
\begin{lstlisting}[frame=single,caption={\VLAD\ example},label=lis:vlad]
(define (f x)
 (let ((n 100003))
  (let outer ((i 1) (y x))
   (if (> i n)
       y
       (outer (+ i 1)
	      (let ((m (l x i)))
	       (let inner ((j 1) (y y))
		(if (> j m)
		    y
		    (inner (+ j 1)
			   (sqrt (* y y)))))))))))
\end{lstlisting}
  \par\vspace*{-2ex}
\end{wrapfigure}
With this, it is possible to reformulate the \Fortran\ example from
Listing~\ref{lis:fortran} in \VLAD\ (Listing~\ref{lis:vlad}).
Then one achieves binomial checkpointing simply by calling
\texttt{($\CHECKPOINTREVERSEJ$ f 3 1)}.

\begin{figure}[tb]
  \centering
  \resizebox{0.95\textwidth}{!}{\begin{tabular}{@{}cc@{}}
      \includegraphics[width=\textwidth]{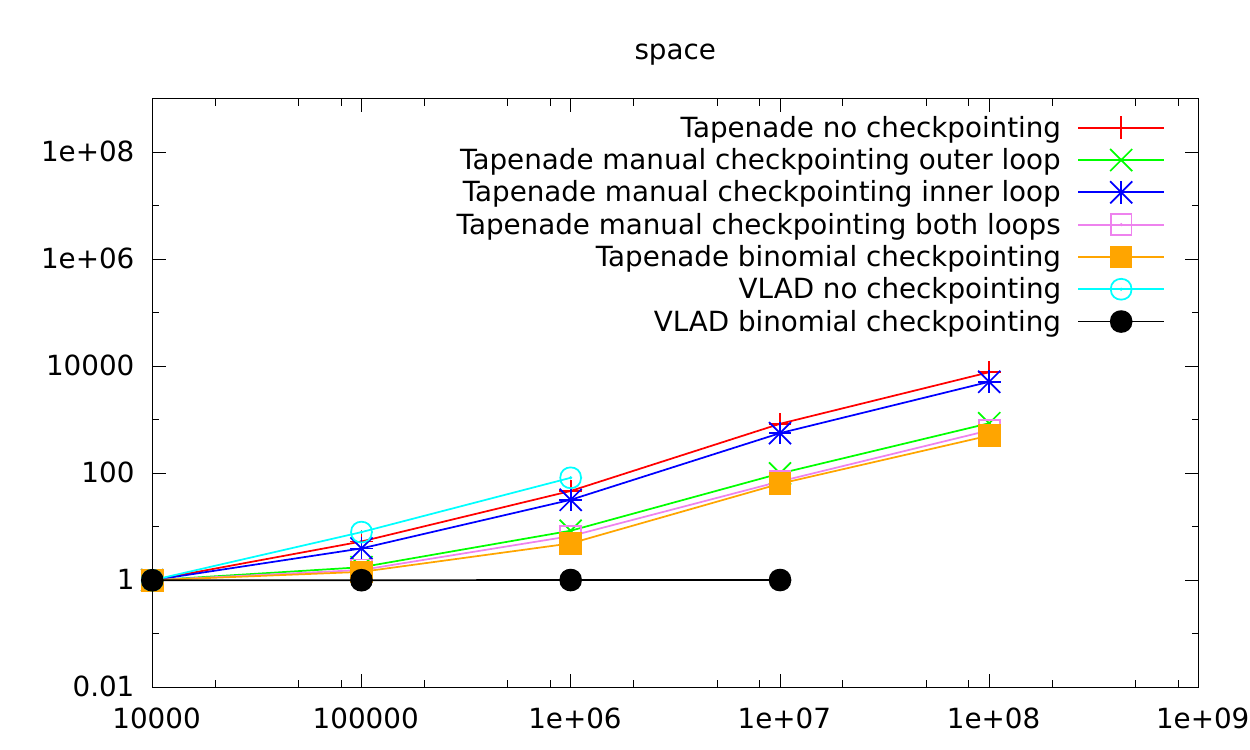}&
      \includegraphics[width=\textwidth]{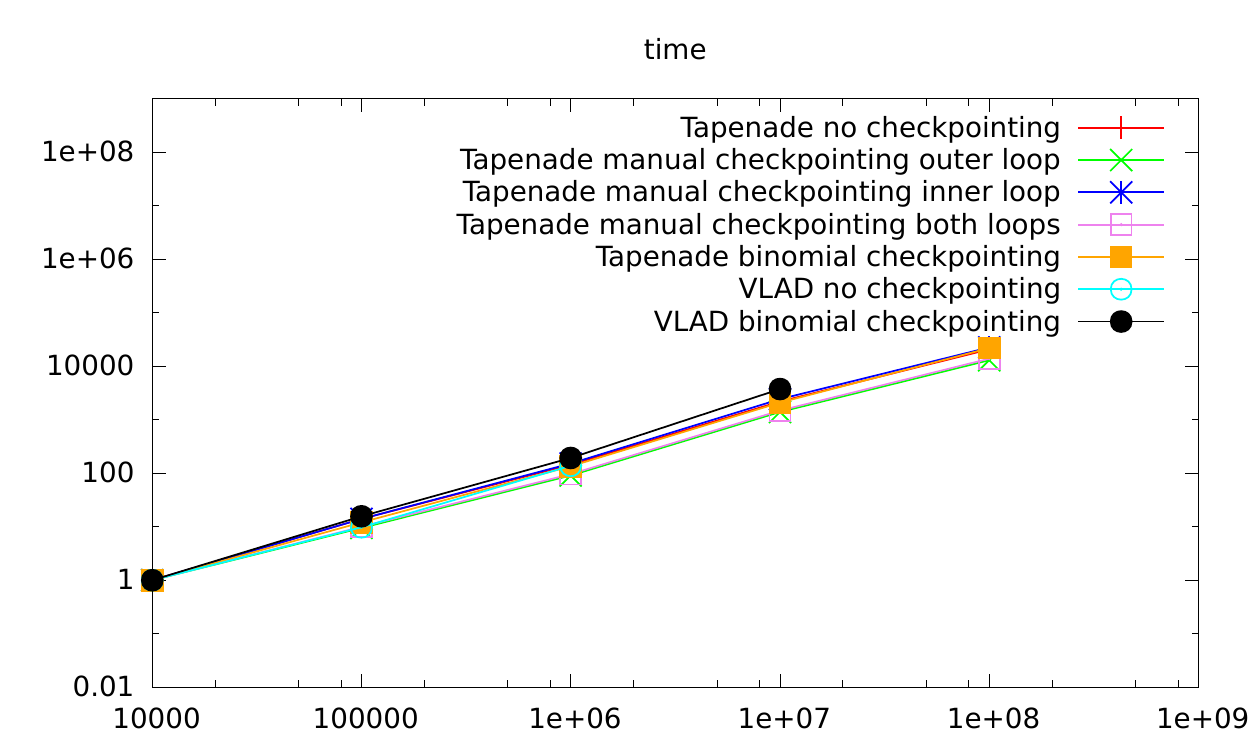}
      \end{tabular}}
  \caption{Space and time usage of reverse-mode AD with various
    checkpointing strategies, relative to the space and time for the first
    datapoint for each respective strategy.}
  \label{fig:plots}
\end{figure}

The efficacy of our method can be seen in the plots (Fig.~\ref{fig:plots}) of
the space and time usage, relative to that for the leftmost datapoint, of the
above \Fortran\ and \VLAD\ examples with varying~$n$.
\Tapenade\ was run without checkpointing, with manual checkpointing only around
the body of the outer loop, with manual checkpointing only around the body of
the inner loop, with manual checkpointing around the bodies of both loops, and
with binomial checkpointing.
\VLAD\ was run with~$\REVERSEJ$ and~$\CHECKPOINTREVERSEJ$.
Note that \Tapenade\ exhibits $O(n)$ space and time usage for all cases,
while \VLAD\ exhibits $O(n)$ space and time usage with~$\REVERSEJ$, but
$O(1)$ space usage and $O(n)$ time usage with~$\CHECKPOINTREVERSEJ$.
The space complexity of~$\CHECKPOINTREVERSEJ$ is the sum of the space required
for the checkpoints and the space required for the tape.
For a general computation of length~$t$ and maximal live storage~$w$, the
former is $O(w\log t)$ while the latter is $O(w)$.
For the code in our example, $t=O(n)$ and $w=O(1)$, leading to the former being
$O(\log n)$ and the latter being $O(1)$.
We observe $O(1)$ space usage since the constant factors of the latter
overpower the former.
The time complexity of~$\CHECKPOINTREVERSEJ$ is the sum of the time required to
(re)compute the primal and the time required to perform the reverse sweep.
For a general computation, the former is $O(t\log t)$ while the latter is
$O(t)$.
For the code in our example, the former is $O(n\log n)$ and the latter is
$O(n)$.
We observe $O(n)$ time usage since, again, the constant factors of the latter
overpower the former.

Other methods present themselves for implementing the general-purpose
checkpointing interface.
One can use \POSIX\ \texttt{fork()} much in the same way that it has been used
to implement the requisite nondeterminism in probabilistic programming
languages like probabilistic \Clang\ \citep{paige2014compilation}.
A copy-on-write implementation of \texttt{fork()}, as is typical, would make
this reasonably efficient and allow it to apply in the host, rather than the
target, and thus could be used to provide an overloaded implementation of
binomial checkpointing in a fashion that was largely transparent to the user.
Alternatively, direct-style code could be compiled into CPS using a CPS
transformation.
A compiler for a language like \VLAD\ can be constructed that generates target
code in CPS that is instrumented with step counting, step limits, and
checkpointing interruptions.
A driver can be wrapped around such code to implement $\CHECKPOINTREVERSEJ$.
Existing high-performance compilers, like \SMLNJ\ \citep{appel2006compiling},
for functional languages like \SML, already generate target code in CPS, so by
adapting such to the purpose of AD with binomial checkpointing, it seems
feasible to achieve high performance.
In fact, the overhead of the requisite instrumentation for step counting, step
limits, and checkpointing interruptions need not be onerous because the
step counting, step limits, and checkpointing interruptions for basic blocks can
be factored, and those for loops can be hoisted, much as is done for the
instrumentation needed to support storage allocation and garbage collection
in implementations like \MLton\ \citep{weeks2006whole}, for languages like
\SML, that achieve very low overhead for automatic storage management.

\section*{Acknowledgments}
This work was supported, in part, by NSF grant 1522954-IIS and by Science
Foundation Ireland grant 09/IN.1/I2637.
Any opinions, findings, and conclusions or recommendations expressed in this
material are those of the authors and do not necessarily reflect the views
of the sponsors.

\bibliographystyle{unsrtnat}
\begin{small}
  \setlength{\bibsep}{0.2ex}
  \bibliography{ad2016a}
\end{small}

\end{document}

%% file: AD.tex
\usepackage{mobi}

\newlength{\arrowbump}
\newlength{\arrowlbump}
\newcommand{\FORWARDPROC}[1]{\overrightarrow{#1}}
\newcommand{\FORWARDPROCx}[1]
{\FORWARDPROC{\rule{0pt}{\arrowbump}\smash[t]{#1}}}

\newcommand{\REVERSEPROC}[1]{\overleftarrow{#1}}
\newcommand{\REVERSEPROCx}[1]
{\REVERSEPROC{\rule{0pt}{\arrowbump}\smash[t]{#1}}}

\newcommand{\IF}{\textbf{if}\;}
\newcommand{\THEN}{\textbf{then}\;}
\newcommand{\ELSE}{\textbf{else}\;}

\newcommand{\LET}{\textbf{let}\;}

\newcommand{\IN}{\textbf{in}\;}

\newcommand{\CLOSURE}[2]{\langle(#1),#2\rangle}

\newcommand{\BUNDLEF}{\rhd}
\newcommand{\BUNDLE}[2]{#1\BUNDLEF#2}
\newcommand{\TAPEF}{\lhd}
\newcommand{\TAPE}[2]{#1\TAPEF#2}

\newcommand{\EVAL}{\mathcal{E}}
\newcommand{\APPLY}{\mathcal{A}}

\newcommand{\JACOBIAN}{\mathcal{J}}

\newcommand{\FORWARDJ}{\FORWARDPROCx{\JACOBIAN}}

\newcommand{\REVERSEJ}{\REVERSEPROCx{\JACOBIAN}}

\newcommand{\defoccur}[1]{\textsl{#1}}


%% file: ad2016a.bbl
\begin{thebibliography}{10}
\providecommand{\natexlab}[1]{#1}
\providecommand{\url}[1]{\texttt{#1}}
\expandafter\ifx\csname urlstyle\endcsname\relax
  \providecommand{\doi}[1]{doi: #1}\else
  \providecommand{\doi}{doi: \begingroup \urlstyle{rm}\Url}\fi

\bibitem[Volin and Ostrovskii(1985)]{Volin1985Aco}
{Yu.}~M. Volin and G.~M. Ostrovskii.
\newblock Automatic computation of derivatives with the use of the multilevel
  differentiating technique --- {I}: {A}lgorithmic basis.
\newblock \emph{Computers and Mathematics with Applications}, 11:\penalty0
  1099--1114, 1985.
\newblock \doi{10.1016/0898-1221(85)90188-9}.

\bibitem[Dauvergne and Hasco{\"e}t(2006)]{Dauvergne2006TDF}
Benjamin Dauvergne and Laurent Hasco{\"e}t.
\newblock The data-flow equations of checkpointing in reverse automatic
  differentiation.
\newblock In Vassil~N. Alexandrov, Geert~Dick van Albada, Peter M.~A. Sloot,
  and Jack Dongarra, editors, \emph{Computational Science -- {ICCS} 2006},
  volume 3994 of \emph{Lecture Notes in Computer Science}, pages 566--573,
  Heidelberg, 2006. Springer.
\newblock ISBN 3-540-34385-7.
\newblock \doi{10.1007/11758549_78}.

\bibitem[Griewank(1992)]{Griewank1992ALG}
Andreas Griewank.
\newblock Achieving logarithmic growth of temporal and spatial complexity in
  reverse automatic differentiation.
\newblock \emph{Optimization Methods and Software}, 1:\penalty0 35--54, 1992.

\bibitem[Griewank and Walther(2000)]{Griewank2000ARA}
Andreas Griewank and Andrea Walther.
\newblock Algorithm 799: {R}evolve: {A}n implementation of checkpoint for the
  reverse or adjoint mode of computational differentiation.
\newblock \emph{ACM Transactions on Mathematical Software}, 26\penalty0
  (1):\penalty0 19--45, mar 2000.
\newblock ISSN 0098-3500.
\newblock \doi{10.1145/347837.347846}.
\newblock Also appeared as Technical University of Dresden, Technical Report
  IOKOMO-04-1997.

\bibitem[Reynolds(1993)]{reynolds1993discoveries}
John~C Reynolds.
\newblock The discoveries of continuations.
\newblock \emph{Lisp and symbolic computation}, 6\penalty0 (3-4):\penalty0
  233--247, 1993.

\bibitem[Bendtsen and Stauning(1996)]{Bendtsen1996FaF}
C.~Bendtsen and Ole Stauning.
\newblock {FADBAD}, a flexible {C++} package for automatic differentiation.
\newblock Technical Report IMM--REP--1996--17, Department of Mathematical
  Modelling, Technical University of Denmark, Lyngby, Denmark, aug 1996.

\bibitem[Sussman and {Steele, Jr.}(1975)]{sussman1975}
Gerald~Jay Sussman and Guy~L. {Steele, Jr.}
\newblock Scheme: An interpreter for extended lambda calculus.
\newblock AI Memo 349, MIT, December 1975.

\bibitem[Paige and Wood(2014)]{paige2014compilation}
Brooks Paige and Frank Wood.
\newblock A compilation target for probabilistic programming languages.
\newblock In \emph{Proceedings of The 31st International Conference on Machine
  Learning}, pages 1935--1943, 2014.

\bibitem[Appel(2006)]{appel2006compiling}
Andrew~W Appel.
\newblock \emph{Compiling with continuations}.
\newblock Cambridge University Press, 2006.

\bibitem[Weeks(2006)]{weeks2006whole}
Stephen Weeks.
\newblock Whole-program compilation in {MLton}, 2006.
\newblock URL
  \url{http://www.mlton.org/References.attachments/060916-mlton.pdf}.
\newblock Workshop on ML.

\end{thebibliography}
